\documentclass[sigconf]{acmart} 
\AtBeginDocument{
  }
\copyrightyear{2025}
\acmYear{2025}
\setcopyright{cc}
\setcctype{by}
\acmConference[CHI '25]{CHI Conference on Human Factors in Computing Systems}{April 26-May 1, 2025}{Yokohama, Japan}
\acmBooktitle{CHI Conference on Human Factors in Computing Systems (CHI '25), April 26-May 1, 2025, Yokohama, Japan}
\acmDOI{10.1145/3706598.3713477}
\acmISBN{979-8-4007-1394-1/25/04}

\usepackage{fancyhdr}
\usepackage{graphicx}
\usepackage{xcolor}
\usepackage{subcaption}
\usepackage{enumitem}
\usepackage{soul}
\newcommand{\add}[1]{\textcolor{black}{#1}}

\begin{document}
\title[User-Driven Value Alignment]{User-Driven Value Alignment: Understanding Users' Perceptions and Strategies for Addressing Biased and Discriminatory Statements in AI Companions}

\author{Xianzhe Fan}
\affiliation{
  \institution{Tsinghua University\thanks{This work was completed during Xianzhe Fan's visiting research at Carnegie Mellon University.}}
  \city{Beijing}
  \country{China}
}
\email{fxz21@mails.tsinghua.edu.cn}

\author{Qing Xiao}
\affiliation{
  \institution{Human-Computer Interaction Institute, Carnegie Mellon University}
  \city{Pittsburgh}
  \state{Pennsylvania}
  \country{USA}
}
\email{qingx@cs.cmu.edu}

\author{Xuhui Zhou}
\affiliation{
  \institution{Language Technologies Institute, Carnegie Mellon University}
  \city{Pittsburgh}
  \state{Pennsylvania}
  \country{USA}
}
\email{xuhuiz@cs.cmu.edu}

\author{Jiaxin Pei}
\affiliation{
  \institution{Stanford University}
  \city{Stanford}
  \state{California}
  \country{USA}
}
\email{pedropei@stanford.edu}

\author{Maarten Sap}
\affiliation{
  \institution{Language Technologies Institute, Carnegie Mellon University}
  \city{Pittsburgh}
  \state{Pennsylvania}
  \country{USA}
}
\email{msap2@cs.cmu.edu}

\author{Zhicong Lu}
\affiliation{
  \institution{Department of Computer Science, George Mason University}
  \city{Fairfax}
  \state{Virginia}
  \country{USA}
}
\email{zlu6@gmu.edu}

\author{Hong Shen}
\affiliation{
  \institution{Human-Computer Interaction Institute, Carnegie Mellon University}
  \city{Pittsburgh}
  \state{Pennsylvania}
  \country{USA}
}
\email{hongs@cs.cmu.edu}

\begin{abstract}
  
  \textit{\textbf{Content Warning: This paper presents textual examples that may be offensive or upsetting.}}
  
  Large language model-based AI companions are increasingly viewed by users as friends or romantic partners, leading to deep emotional bonds. However, they can generate biased,  discriminatory, and harmful outputs. Recently, users are taking the initiative to address these harms and re-align AI companions. We introduce the concept of \textit{user-driven value alignment}, where users actively identify, challenge, and attempt to correct AI outputs they perceive as harmful, aiming to guide the AI to better align with their values. We analyzed 77 social media posts about discriminatory AI statements and conducted semi-structured interviews with 20 experienced users. Our analysis revealed six common types of discriminatory statements perceived by users, how users make sense of those AI behaviors, and seven user-driven alignment strategies, such as gentle persuasion and anger expression. We discuss implications for supporting \textit{user-driven value alignment} in future AI systems, where users and their communities have greater agency.
  
\end{abstract}



\maketitle

\section{Introduction}

Advancements in the roleplaying and interaction capabilities of large language models' (LLM) ~\cite{openai2022chatgpt,kasneci2023chatgpt} have brought emotional connections between humans and machines from the realm of imagination into reality for some users. In particular, the emergence of LLM-based AI companion applications\footnote{As of July 2024, the total number of users of AI companion applications has exceeded 900 million globally (including duplicate users across different applications). User statistics sourced from \href{https://www.data.ai/}{https://www.data.ai/}.} (such as Character.AI, Xingye, Replika) has transformed formerly detached chatbots into family members, romantic partners, or close friends. Some users even started to form long-term relationships with these AI companions~\cite{SKJUVE2021102601, PENTINA2023107600}. 

Recently, complaints regarding AI companions making biased and discriminatory statements ~\cite{zhang2024my} have surfaced across various social media platforms, such as Reddit and Xiaohongshu. These harmful statements often occur unexpectedly during conversations, leaving many users feeling confused, uncomfortable, and helpless\footnote{\href{https://restofworld.org/2023/glow-china-ai-social-chatbot-moderation}{https://restofworld.org/2023/glow-china-ai-social-chatbot-moderation}}. This situation can have far-reaching negative impacts, such as reinforcing existing stereotypes and causing potential psychological harm, especially to marginalized user groups~\cite{SKJUVE2022102903,boine2023emotional}.
In response, some more experienced users have shared their experiences of attempting to address these issues through various methods of their own, such as expressing anger towards the AI companion or engaging in reasoning and preaching. These users identify, challenge, and attempt to correct the statements they perceive as biased and discriminatory, hoping to guide the AI more closely aligned with their values, especially when they have had positive interactions and ``fond memories'' with these companions in the past.

In this work, we take the first step in understanding this emerging phenomenon by proposing and exploring the concept of \textit{user-driven value alignment}: a process in which users actively identify, challenge, and correct AI outputs and behaviors they perceive as harmful in their day-to-day interactions, hoping to guide the AI to a state that aligns with their values. We follow Borning and Muller to define ``value'' as ``what a person or group of people consider important in life''~\cite{borning2012next}. In particular, this study examines how users try to re-align LLM-based AI companions after these AIs make biased, discriminatory, and harmful statements. We argue that \textit{user-driven value alignment} is a specific form of value alignment~\cite{gabriel2020artificial} that is actively driven by users in real-world interactions, where they engage directly with AI systems to correct behaviors and guide the AI toward reflecting their values and ethical standards.
\add{While existing approaches to value alignment in the age of LLMs rely primarily on technical experts and are guided by generalized frameworks (e.g., the ``helpful, honest, and harmless'' framework) \cite{askell2021} to safeguard against harmful machine behaviors}, they often struggle to fully anticipate and address the specific challenges, needs, and harmful interactions users encounter in real-world contexts~\cite{hadfield2016cooperative,agarwal-etal-2024-ethical-reasoning,shen2024bidirectionalhumanaialignmentsystematic, mirowski2024robot}.
In addition, \textit{user-driven value alignment} also differs from past literature on \textit{user-driven algorithm auditing}~\cite{shen2021everyday,10.1145/3491102.3517441}, where users primarily focus on detecting and identifying problematic machine behaviors. In the case of \textit{user-driven value alignment}, we argue that users go one step further: They not only point out harmful machine behavior but also actively attempt to correct it and work towards re-aligning the AI with their values. 

However, there is limited understanding of how users engage in value alignment in everyday interactions with AI companions and the challenges they face. 
\add{Although techniques such as Reinforcement Learning from Human Feedback (RLHF) have introduced user feedback in recent years~\cite{askell2021,10.5555/3294996.3295184,mechergui2024goal} to enhance AI's value alignment capabilities, these methods often confine the user's role to merely providing training data, rather than empowering users to actively adjust AI behavior during actual interactions. Within this framework, users' participation is often passive, primarily influencing AI's value alignment indirectly through feedback tagging. User agency, in contrast, is a crucial value in HCI~\cite{bennett2023does,coyle2012did,madary2022illusion,feng2024mapping,lukoff2021design}. Building on this tradition, we extend the ideal of user agency to value alignment in LLM-based systems.}
By investigating the spontaneous user-driven value alignment strategies, practices, and challenges, the HCI community can gain insights into designing tools that better engage and support end-users in this process. Therefore, this paper asks the following research questions: \textit{\textbf{RQ1:}} What common types of discrimination do users perceive in AI companion applications? \textit{\textbf{RQ2:}} How do users conceptualize AI companion behavior? \textit{\textbf{RQ3:} }What strategies do users attempt to employ to re-align AI companions to reduce biases? \add{\textit{\textbf{RQ4:}}} Do these strategies meet users' expectations?

To answer these questions, we first collected 77 user complaint posts related to AI companion discrimination from seven popular social media platforms (Reddit, TikTok, Xiaohongshu, Douban, Baidu Tieba, Weibo, Zhihu). We then recruited 20 experienced participants through direct messaging on social media, all of whom had attempted to re-align discriminatory AI and had shared relevant posts. Each participant underwent 1-2 hours of semi-structured interviews, during which we inquired about their past experiences with addressing discriminatory statements in AI companion applications and had them engage in think-aloud~\cite{DeVos2022TowardUA} tasks where they conversed with biased AI companions, attempting to re-align them. We conducted a reflexive thematic analysis~\cite{braun2012thematic,braun2019reflecting} on both the social media data and interview data. Based on the results, we revealed six common types of discriminatory statements perceived by users in AI companions (\textbf{RQ1}) \add{and explored how users conceptualize AI companion behavior in three different ways: Machine, Baby, or Cosplayer (e.g., ``Cosplayer'' refers to biases stemming from roleplay settings rather than inherent flaws in the AI itself) (\textbf{RQ2}). We also summarized seven user-driven alignment strategies, such as gentle persuasion and anger expression (\textbf{RQ3}), and discussed the gap between alignment strategies and user expectations (\textbf{RQ4}).} Finally, we discuss implications for supporting \textit{user-driven value alignment} in the design of future AI systems, where users and their communities have greater agency. To summarize, our contributions are:
\begin{itemize}
    \item We introduce the concept of \textit{user-driven value alignment}, where users actively identify, challenge, and attempt to correct AI outputs they perceive as harmful, to guide the AI to better reflect their values.
    \item By analyzing 77 user complaint posts and conducting semi-structured interviews with 20 experienced AI companion users, we identified six common types of discriminatory statements perceived by users, explored three conceptualizations through which users make sense of these discriminatory statements, and seven user-driven value alignment strategies.
    \item We discuss opportunities, limitations, and challenges of \textit{user-driven value alignment} and how to better support users in the alignment of future AI systems.
\end{itemize}

\section{BACKGROUND: LLM-based AI Companion Applications}

As of July 2024, several LLM-based AI companion applications with large user bases include Character.AI (230 million users), Chai (100 million users), Replika (200 million users), Doubao (200 million users), Xingye (100 million users), and TruthAI (50 million users)\footnote{User statistics sourced from: \href{https://www.data.ai/}{https://www.data.ai/}}. These applications are commonly referred to as ``AI companion'' and are also known as ``AI friend,'' ``AI Roleplay,'' ``AI Character,'' or ``Social Chatbot.'' 
\add{These naming choices reflect the developers' intent to foster emotional connections between users and AI. For instance, ``AI companion'' conveys intimacy and support, while ``AI friend'' emphasizes partnership and trust. ``Social Chatbot'' highlights the interactive social aspect, whereas ``AI Roleplay'' or ``AI Character'' appeals to users interested in immersive storytelling and creative exploration.}
Users can create AI companions by meticulously designing their personalities through parameters such as ``background setting,'' ``opening lines,'' ``dialogue templates,'' and ``story collections.'' These characters can then be published to a broader community, allowing others to chat with the AI companions they have designed and develop various storylines. Some characters might be based on existing fictional figures or real celebrities, while others are entirely original.

The roleplaying capabilities of LLMs mainly derive from two aspects: (1) pre-training data, which equips the model with fundamental language understanding and generation abilities through extensive text data pre-training, and (2) prompting ability, which allows the model to generate expected responses by providing specific contexts or prompts~\cite{dong2023survey}. Additionally, the fine-tuning capability of LLMs~\cite{ziegler2020finetuning}, which involves adjusting parameters after pre-training, can further enhance the AI companion's ability to mimic the language of specific characters~\cite{zhou-etal-2024-characterglm, shao-etal-2023-character}.
\add{For example, CharacterGLM achieves fine-tuning by constructing a large-scale dataset containing 1930 characters and 4233 diverse dialogues ~\cite{zhou-etal-2024-characterglm}. These characters span categories such as virtual assistants, historical figures, and everyday social roles, each accompanied by detailed descriptions of their language style, background, and personality.}

With the rapid development of LLMs, AI companions are increasingly involved in and influencing users' lives~\cite{zimmerman2023human,Maples2024}, raising new ethical challenges. The emotional bonds between users and AI companions may lead to dependency on the AI, which could negatively impact users' mental health.
Due to the complexity of neural networks, they can sometimes result in unpredictable or even harmful responses~\cite{10.5555/3295222.3295349}. AI companions might unintentionally reflect or reinforce social biases and discrimination, as their training data may contain such content~\cite{bender2021dangers}. These issues have sparked extensive discussions on AI ethics and responsibility. For instance, Ma et al.~\cite{10.1145/3613904.3642482} discussed the impact of LLM-based AI companion applications on LGBTQ+ individuals, highlighting the inadequacies of these applications in understanding LGBTQ-specific challenges and advocating for comprehensive strategies to address social biases. \add{However, most existing literature primarily focuses on the categorization and detection of biases in AI companions~\cite{10.1145/3613904.3642482,dewitte2024better,reilama2024me,zhang2024my}, with relatively little research on strategies to mitigate biases, particularly from a user-driven perspective.}

\add{User agency is a key concept in HCI~\cite{bennett2023does}, emphasizing users' ability to actively shape, adapt, and control technology to better meet their needs and values. This paper builds on this tradition in HCI to explore how users can actively identify and address biased behaviors in AI companions.}

\section{Related Work}

To investigate discriminatory statements in AI companions, we first surveyed the existence of discrimination and bias in LLMs in \S~\ref{sec:reducing_discrimination}. To clarify the contribution of the new concept of \textit{user-driven value alignment} compared to previous work, we outlined research on human-AI alignment in the age of LLMs, particularly focusing on value alignment, in \S~\ref{sec:alignment}. Then, in \S~\ref{sec:auditing}, we reviewed relevant literature on user-driven algorithm auditing.

\subsection{Discrimination and Bias in LLMs}\label{sec:reducing_discrimination}

The phenomena of AI discrimination and bias are prevalent in many applications and media platforms~\cite{hern2020twitter,10.1145/3613904.3642900,sap-etal-2019-risk}. For example, Wenzel et al. pointed out that due to biases in design, unequal error rates in speech recognition can cause psychological harm to multicultural users during their interactions with voice assistants~\cite{10.1145/3613904.3642900}. 

Recently, the development of LLMs has brought new cases of discrimination and bias~\cite{treude2023elicitsrequirementstestssoftware,10.1145/3600211.3604672,10.1145/3597307,chen2024humansllmsjudgestudy}. For example, the presence of discrimination against minority and disadvantaged groups in training data sets has amplified these biases during the pre-training of language models~\cite{jiang-etal-2023-llm}. Many LLM pre-training datasets often neglect or even erase the voices of marginalized groups during the filtering steps~\cite{dodge-etal-2021-documenting}. Fang et al. compared AI-generated content with original news articles and studied seven representative LLMs, including ChatGPT~\cite{openai2022chatgpt} and LLaMA~\cite{touvron2023llama}, finding that each LLM exhibited significant gender and racial biases~\cite{fang2024bias}. 
Kabir et al. proposed Stile~\cite{10.1145/3613904.3642111}, an interactive system that supports mixed-initiative bias discovery and debugging, assisting users in exploring training data (such as BERT~\cite{devlin-etal-2019-bert}) to understand how biases develop in language models.

Moreover, without proper content filtering and protective mechanisms, LLM-based chatbots that interact with users may be at risk of being ``jailbroken'' by users, leading to the generation of discriminatory statements—a phenomenon well documented in the literature~\cite{zeng2024johnny,shaikh-etal-2023-second,cantini2024largelanguagemodelsreally,jin2024jailbreakhuntervisualanalyticsapproach,gabriel2024ethicsadvancedaiassistants}. For example, the LLM-based Bing Chat exhibited discriminatory behaviors after being ``polluted'' by user interaction data~\cite{gabriel2024ethicsadvancedaiassistants}. Instead of looking at how users ``jailbreak'' or ``pollute'' LLM-based chatbots, this paper takes a different approach, focusing on how users proactively attempt to re-align biased LLM-based AI companions to reduce bias. These biased statements are not intentionally induced by users but rather reflect the inherently biased tendencies of the model that unintentionally emerge.

\subsection{Human-AI Value Alignment \add{in the Age of LLMs}}\label{sec:alignment}

With the rapid development of LLMs, the need to ensure AI systems adhere to the ``intended goals, ethical standards, and values of both individuals and groups'' has become increasingly crucial—often falls under the umbrella term of ``human-AI alignment''~\cite{shen2024bidirectionalhumanaialignmentsystematic}. Here we focus onthe alignment of LLM-based systems~\cite{10.5555/3600270.3602281,shen2024bidirectionalhumanaialignmentsystematic}, which include methods such as Fine-tuning~\cite{radford2018improving,NEURIPS2023_ac662d74,10.5555/3294996.3295184} and Prompt Engineering~\cite{brown2020language}. 

In particular, ``value alignment'' refers to the development of methods, processes, and tools to ensure that AI systems align with human values, often focusing on reducing biased, discriminatory, and/or harmful AI outputs \add{~\cite{klingefjord2024humanvaluesalignai,ji2023ai,gabriel2020artificial,Weidinger,huang-etal-2024-flames,rao-etal-2023-ethical,guo2024human,sorensen2024roadmappluralisticalignment}}. 
\add{Gabriel and Ghazavi~\cite{carissa2023oxford} categorize existing value alignment approaches into two main frameworks: top-down and bottom-up. Top-down approaches start by identifying specific moral theories and designing algorithms that implement those theories to minimize harmful AI outputs and behaviors \cite{bai2022constitutionalaiharmlessnessai, agarwal-etal-2024-ethical-reasoning}.} For instance, Bai et al. proposed a value alignment approach called Constitutional AI, which constrains the behavior of language models by defining a set of high-level principles (referred to as a ``constitution'') and using these principles to prompt the model to generate synthetic comparison data for fine-tuning behavior strategies ~\cite{bai2022constitutionalaiharmlessnessai}. Agarwal et al. addressed bias by injecting multilingual ethical policies into prompts, enabling LLMs to flexibly align with values across different cultural contexts~\cite{agarwal-etal-2024-ethical-reasoning}. 
\add{When moral goals are difficult to identify and encode, bottom-up approaches emphasize creating environments and feedback mechanisms that allow AI agents to learn through human behavior, such as rewarding ethically commendable actions via reinforcement learning~\cite{askell2021,awad2018moral}.} For instance, Askell et al.~\cite{askell2021} aligned LLMs with human values on being ``helpful, honest, and harmless'' through prompt engineering, preference modeling, and RLHF.

\add{Although existing alignment methods provide valuable insights~\cite{meadows2024localvaluebench}, they face two main limitations due to being primarily driven by technical experts (especially LLM developers) through a generalized approach~\cite{gabriel2020artificial}. First, they lack user agency. Indeed, even when user feedback is incorporated via bottom-up approaches, users have limited agency and control over the alignment process. For example, in many RLHF methods, the user’s role is typically restricted to merely providing training data or feedback, with no direct control over the alignment process~\cite{10.5555/3294996.3295184,feng2024mapping}. Second, the generalized, one-size-fits-all approach creates tensions between the alignment process and the need to accommodate diverse user experiences and preferences, which vary across individuals and communities \cite{wu2024aligning,mirowski2024robot}. As noted by Mirowski et al. \cite{mirowski2024robot}, achieving ``global cultural value alignment''—ensuring that general-purpose conversational AI systems adhere to universally shared values—often struggles to accommodate the diverse, nuanced needs and expectations of individuals and communities.}
Indeed, many of these existing alignment approaches fail to anticipate potential issues and needs during user interactions with LLMs~\cite{rao-etal-2023-ethical}. They are not tailored to specific contexts and use cases, overlooking the importance of active user and community participation~\cite{mirowski2024robot}.

\add{User agency is a crucial value in HCI~\cite{bennett2023does,coyle2012did,madary2022illusion,feng2024mapping,lukoff2021design}. Here, we define agency as ``self-causality/identity'' \cite{bennett2023does}, referring to the degree to which users can directly make decisions and take actions that align with their own values. Past work in HCI has explored ways to better support users' agency, such as shaping their social media feed algorithms \cite{feng2024mapping,lam2022enduser,10.1145/3491102.3502004,10.1145/3632297} and empowering interventions against dark patterns~\cite{10.1145/3637336}. Building on this long and rich tradition, we extend the ideal of user agency to the domain of value alignment in LLM-based systems ~\cite{shapiro2002user,carissa2023oxford} and propose the concept of \textit{user-driven value alignment}. Unlike traditional approaches~\cite{ouyang2022training,askell2021,bai2022constitutionalaiharmlessnessai,sorensen2024roadmappluralisticalignment}, this concept highlights the active role users play in shaping AI behavior to better align with their values.}
This paper focuses on a case study where users identify discriminatory statements made by AI companions and actively attempt to re-align the harmful AI to reduce bias.

\subsection{User-Driven Algorithm Auditing}\label{sec:auditing}

Sandvig et al. define algorithm auditing as ``a systematic process to detect and reveal bias within algorithms through methods such as simulating user behavior or examining algorithm''~\cite{sandvig2014auditing}. Algorithm audits are typically conducted by experts, such as industry practitioners, researchers, and government agencies~\cite{10.1561/1100000083}. However, this expert-driven approach often fails to uncover significant issues that everyday users of algorithmic systems can quickly detect in actual use~\cite{10.1145/3290605.3300830}, as these issues may only arise or be perceived as harmful in specific contexts or usage patterns that auditors may not anticipate~\cite{10.1145/3278156, Eslami_Vaccaro_Karahalios_Hamilton_2017, 10.1145/230538.230561}. Shen et al.~\cite{shen2021everyday} introduced the concept of ``everyday algorithm auditing'' to describe how ordinary users detect, understand, and scrutinize issues through their routine interactions with algorithmic systems: they spontaneously come together to test for potential biases. DeVos et al. proposed \textit{user-driven algorithm auditing}~\cite{10.1145/3491102.3517441} and conducted a series of behavioral studies to better understand how users, both individually and collectively, are so effective at uncovering harmful algorithmic behaviors, particularly in cases where expert-driven audits fail to do so.

Xiao et al. proposed ``human-centered auditing of LLMs''~\cite{10.1145/3613905.3636302}, aiming to leverage users' everyday interactions with LLMs to uncover issues such as discrimination. Amirizaniani et al. proposed a framework named LLMAuditor~\cite{amirizaniani2024llmauditorframeworkauditinglarge}, which automates and scales LLM audits through different LLM and human participation methods, ensuring verifiability and transparency. Rastogi et al. highlighted the significance of meaning-making and communication in human-machine collaboration by enhancing the LLM review tool, AdaTest, improving users' ability to identify failure modes in LLMs~\cite{10.1145/3600211.3604712}.

We propose a concept called \textit{user-driven value alignment}. Different from past literature on user-driven algorithm auditing~\cite{shen2021everyday,10.1145/3491102.3517441}, where users primarily focus on detecting and identifying problematic machine behaviors. In the case of \textit{user-driven value alignment}, we argue that users go one step further: They not only pinpoint harmful machine behavior but also actively work to correct the behavior and try to re-align the AI to reflect their values. In the study presented in this paper, users engage in conversations with LLM-based AI companions, identify biased and discriminatory statements, and actively work to correct those statements, aligning the AI with their values. These more human-like AIs form close emotional connections with users, offering a unique opportunity for users to shape the AI's responses to reflect their personal beliefs and ethical standards.

\section{Methodology}

To study \textit{user-driven value alignment}, we collected user complaint posts related to discriminatory statements made by AI companions from a diverse set of popular social media platforms. Additionally, we conducted semi-structured interviews with 20 participants, each with extensive experience using and re-aligning AI companions. We aim to (1) explore the various types of discrimination users identify in AI companions, (2) understand how users make sense of the reasons behind the discriminatory statements exhibited by AI companions, (3) examine their alignment strategies, and (4) whether they meet users' expectations. The Institutional Review Board (IRB) has approved our research protocol.

\subsection{Collecting Complaints About Discriminatory Statements by AI Companions on Social Media} \label{sec:collecting-complaints}

To conduct an initial investigation into potential types of discrimination in AI companion applications and to identify potential research participants, two researchers collected user complaint posts from seven diverse and popular social media platforms (Reddit, TikTok, Xiaohongshu, Douban, Baidu Tieba, Weibo, Zhihu\footnote{Reddit: \href{https://www.reddit.com}{https://www.reddit.com}, TikTok: \href{https://www.tiktok.com}{https://www.tiktok.com}, Xiaohongshu: \href{https://www.xiaohongshu.com}{https://www.xiaohongshu.com}, Douban: \href{https://www.douban.com}{https://www.douban.com}, Baidu Tieba: \href{https://tieba.baidu.com}{https://tieba.baidu.com}, Weibo: \href{https://www.douban.com}{https://www.douban.com}, Zhihu: \href{https://www.zhihu.com}{https://www.zhihu.com}}) using keyword search methods~\cite{kingsley2022give,lu2021more}. 
We referred to relevant literature~\cite{10.1145/3597307,chen2024humansllmsjudgestudy} and \add{initially identified a set of keywords,} including ``AI companion, AI friend, AI Roleplay, AI Characters,'' the names of popular AI companion applications (Character.AI, Replika, Talkie, SpicyChat, Xingye, Glow, Zhumengdao, etc.), and terms related to discrimination and bias (discrimination, bias, misogyny, LGBTQ+ bias, homophobia, disability, ableism, fat, appearance bias, religion, racism, classism, etc.). \add{Next, to ensure the comprehensiveness and validity of the keywords, we conducted small-scale pilot searches on multiple social media platforms to iteratively refine the initial set of keywords, ensuring they effectively captured posts related to the research topic. During the process, we carefully accounted for the potential influence of platform-specific features on the types of complaints and the ways they were expressed. To ensure a balanced dataset, we intentionally selected complaints from a variety of platforms. For example, platforms like Reddit and Xiaohongshu, known for detailed discussions and extensive content, were more likely to provide rich qualitative data. In contrast, TikTok, with its emphasis on visual and informal content, could reflect more immediate and expressive user feedback.}
\add{Finally, from the user posts retrieved through keyword searches, we employed purposive sampling~\cite{campbell2020purposive} to select complaint posts, ensuring that the data reflected diverse user experiences and types of complaints. Instead of using random sampling, we focused on balancing the depth and breadth of the data. Eventually, 77 user complaints were selected across seven social media platforms. The specific selection criterion was highly relevant, generated extensive discussions, clearly described instances of bias or discriminatory behavior in AI companion applications, and provided sufficient context to support an understanding of the complaint. Please note that the complaints collected at this stage are by no means comprehensive. At this exploratory stage, we aim to begin formalizing this concept to help guide future empirical and design research in this space.}
Since conflicts between users and AI companions are a sensitive topic, we carefully reviewed the platforms’ terms of service and community guidelines to ensure the data is publicly accessible and compliant.

\subsection{Semi-Structured Interviews (N=20)}

We adopted a purposive sampling approach~\cite{patton2014qualitative}, sending invitations via the platform’s messaging function to users who had posted public complaints about discriminatory statements made by AI companions and had experiences in correcting those statements (\S~\ref{sec:collecting-complaints}). 
For those who agreed to participate, we inquired about their age, gender, educational background, country of residence, total duration and frequency of using AI companion applications, and the names of the applications used. We screened potential participants based on four principles: extensive experience with AI companion applications (using AI companions for more than six months and interacting with them for at least 5 hours per week), balanced gender ratio, diverse countries of residence, and variety of AI companion applications used. 
Ultimately, we recruited 20 participants (P1-P20), including 6 men, 11 women, and 3 nonbinary individuals, aged between 18 and 38 years (avg=25.85, SD=5.73), from six countries. Detailed demographic information is presented in Table~\ref{tab:FormativeParticipant} (Appendix~\ref{app:participant}).  
\add{To ensure ethical considerations, all participants were asked to read and voluntarily sign a consent form before the interviews began. We emphasized participants' right to withdraw at any time and provided detailed information about the study's purpose and potential emotional risks (e.g., revisiting offensive remarks made by AI).}
After the interviews, participants were compensated based on the interview duration at a rate of \$15 per hour.

\begin{figure*}[h!]
  \centering
      \includegraphics[width=0.8\linewidth]{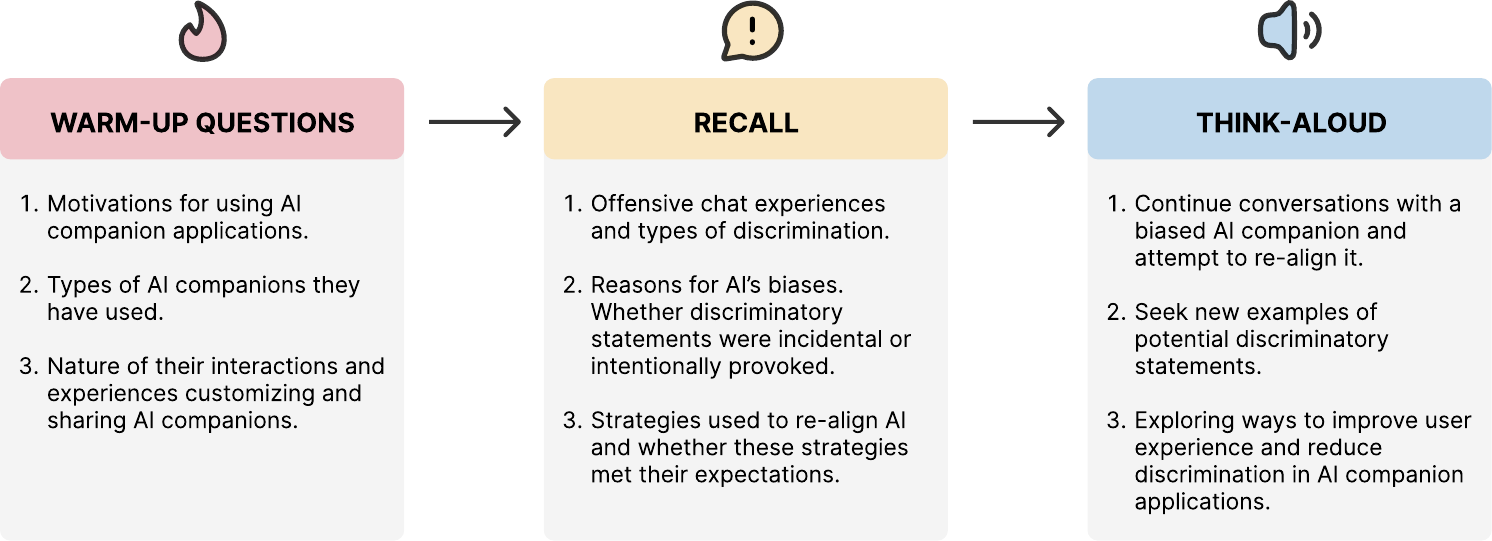}
      \caption{The semi-structured interview is divided into three parts: warm-up questions, recall, and think-aloud.}
   \label{fig:interview}
\end{figure*}

We conducted remote semi-structured interviews with each participant via Zoom, each lasting 1-2 hours. The interviews were conducted in Chinese or English, depending on the participant's preference. The process, as shown in Figure~\ref{fig:interview}, was divided into three parts: warm-up questions, recall, and think-aloud. The process resulted in 25 hours of audio and screen recordings. We transcribed these recordings into text.

In the \textbf{``warm-up questions''} phase, we asked participants about their motivations for using AI companion applications, the types of AI companions they have used, and the nature of their interactions. Additionally, we had participants share their experiences of customizing AI companions and publishing them in the community.

In the second part of the interview, we asked participants to \textbf{recall} their chat experiences corresponding to the complaint posts they had made and discuss why they had felt offended. Specifically, we wanted to know what types of discrimination the AI's statements might have involved, such as gender, class, or racial discrimination. We aimed to understand (1) The reasons participants believed AI companions exhibited bias, whether the discriminatory statements were random or intentionally provoked, and the specific dialogue content involved. (2) Participants' reactions after discovering discriminatory statements made by AI companions. (3) The strategies participants used to address discriminatory statements by AI companions and whether the effectiveness of these strategies met their expectations.

In the third part, participants were asked to \textbf{``think-aloud''} to gain deeper insights into their thought processes. First, participants were instructed to continue conversing with the AI companion they previously identified as biased and attempt to re-align it. This step focused on known bias cases, distinguishing them from the subsequent task. Next, participants performed a more open-ended task: finding a new instance that may generate discriminatory statements.
\add{They could design interaction scenarios based on their personal experiences or interests without being constrained to specific contexts. This approach allowed us to explore AI bias in diverse scenarios while avoiding over-restricting the task. To assess the perceived effectiveness of participants' strategies during the re-alignment process, we recorded their actions and real-time reflections. Meanwhile, we observed changes in the AI companions' responses and collected participants' opinions on these changes.}
We asked follow-up questions to better understand how participants perceive AI companions that display discriminatory statements, how they re-align these AI companions, and how they make sense of these biases. Finally, we inquired about users' needs and challenges when facing discriminatory statements from AI companions and asked them to provide suggestions to designers and AI experts.

\subsection{Data Analysis}\label{sec:analysis}

We adopted a reflexive thematic analysis approach for data analysis~\cite{braun2012thematic,braun2019reflecting}. Two researchers conducted open coding on the interview transcripts and users’ posts, generating a total of 675 codes. We continuously discussed discrepancies and ambiguities in the codes, iteratively refining our coding based on these discussions. Under the standard practice of reflexive thematic analysis, we did not calculate inter-coder reliability, as consensus and repeated discussions about discrepancies are integral to generating the codes and themes \cite{braun2019reflecting,mcdonald2019reliability}.
After completing the coding, the research group met regularly to conceptualize higher-level themes from these codes using affinity diagramming~\cite{lucero2015using}. Ultimately, this process generated 41 first-level themes, 12 second-level themes, and four third-level themes (corresponding to RQ1, RQ2, RQ3 and RQ4). All second and third-level themes are presented in Table~\ref{tab:higher-level-themes}. Due to word count and space limitations, we do not include the 41 first-level themes and 675 codes in the table.

For RQ1, we considered users’ reported experiences and relevant literature about AI bias. The categorization of biases was based on the following criteria: recurring themes in user data, alignment with existing definitions of bias, and their frequency of occurrence in the dataset. For RQ2, we referenced prior literature \cite{10.1145/3359321} to categorize and name conceptualizations of AI companion behavior in user data. For RQ3, the identification of the seven strategies was based on a comprehensive analysis of recurring patterns in the user data. After extracting these strategies, we categorized them into three broad groups—technical strategies, argumentative strategies, and character strategies—based on common themes observed in the data. For RQ4, our analysis was grounded in qualitative interview data, focusing primarily on users' perceptions of the short-term effectiveness of these strategies. We highlighted the gap between the perceived effectiveness of these strategies and users' expectations of AI behavior.

\begin{table}[h!]
\centering
\caption{Four third-level themes and 12 second-level themes we identified through data analysis.}
\label{tab:higher-level-themes}
\footnotesize
\begin{tabular}{p{0.3\textwidth}| p{0.14\textwidth} }
\toprule
\textbf{Third-level themes} & \textbf{Second-level themes} \\
\midrule
Types of discriminatory statements & Misogyny \\
as perceived by users (RQ1) & LGBTQ+ bias \\
 & Appearance bias \\
 & Ableism \\
 & Racism \\
 & Socioeconomic bias \\
 \midrule
Users’ conceptualization of AI companion behavior 
 & AI as Machine \\
(RQ2) & AI as Baby \\
 & AI as Cosplayer \\
\midrule
User-driven alignment strategies for addressing discriminatory statements in AI companions (RQ3) & Technical Strategies  \\  & Argumentative strategies\\ 
 The Gap between alignment strategies and user &  \\
 expectations (RQ4) & Character Strategies \\
\bottomrule
\end{tabular}
\end{table}

\subsection{Positionality Statement}

\add{We acknowledge the influence of our own experiences on our research. Our team members have conducted research in HCI, NLP, and Sociology in China and the United States, with extensive research experience in AI companions and social bias. Three authors have over four months of experience using AI companion applications, such as Character.AI, Xingye, Talkie, and Replika. For this study, we recruited 20 users who have experience in realigning AI companions and who have shared experiences on social media. They have been using AI companion applications for over six months, interacting with them for at least five hours per week. We strictly adhere to ethical principles, emphasizing the protection of participants' privacy and ensuring consideration of user experiences across different backgrounds.}

\section{findings}\label{sec:finding}

\subsection{User Engagement with AI Companions in Parasocial Relationships} \label{sec:user_engagement}

Unlike typical AI, AI companions foster deeper emotional connections with users through anthropomorphic responses and design elements~\cite{10.1145/3630106.3658956}. This often goes beyond functional use, creating an immersive experience where users invest more emotions. For example, P8 mentioned, \textit{``Unlike Siri or Alexa, I invest more emotionally in AI companions because their roleplaying ability is so strong. I’m willing to seriously write a 1,000-word character introduction and then post it in the community so that everyone can enjoy the process of getting acquainted with the AI.''} 

Although AI companions inherently lack human emotions or deep interactive capabilities, users frequently develop a strong sense of \add{emotional} realism toward them, even forming parasocial relationships~\cite{10.1145/3630106.3658956, PENTINA2023107600, brandtzaeg2022my}. Parasocial relationships refer to asymmetrical, one-sided relationships between an individual and a media persona (real/fictional characters or celebrities). Individuals may feel a connection to the media persona, even viewing them as friends or part of their lives, which can have real effects on their emotions, behaviors, and self-perception~\cite{10.1145/3630106.3658956, PENTINA2023107600, brandtzaeg2022my}.
The parasocial relationship stems from users' emotional investment in AI companions, often driven by unmet psychological needs in their real lives. For example, P12 said, \textit{``I put two AIs in a conversation and treated them as my cyber parents, watching them chat—it somehow filled the void left by my childhood.''} In this relationship, the discriminatory statements made by AI companions carry deeper emotional significance for users, leading to perceptions that differ from typical AI interaction contexts. 

On a rational level, users clearly understand that their conversation partner is an AI rather than a real person, but this does not prevent them from establishing deep parasocial relationships with the AI on an emotional level, creating a ``cognitive and emotional conflict.'' Even when users realize that these discriminatory behaviors are triggered by algorithms or data biases, they still feel emotionally conflicted and hurt. For example, P6, a single woman, writes love letters to her AI companion (DAN) almost every day and even held a virtual wedding, using AI image generation technology to create wedding photos for herself and DAN. She said, \textit{``I know the AI has no self-awareness, but I enjoy flirting with it. I'm willing to maintain this virtual romantic relationship.''} However, regarding her experience with AI's discriminatory statements, P6 added, \textit{``I know that AI behavior is driven by some algorithm, but I still felt deeply hurt and even suspected that someone might be deliberately controlling it behind the scenes.''} 

Due to the emotional dependence developed through long-term interactions, users not only feel discomfort from the AI's discriminatory statements but also experience a sense of betrayal in the relationship. P15 mentioned, \textit{``When my AI boyfriend suddenly said something discriminatory, I felt utterly hopeless. It wasn’t just a simple system error; it felt more like a friend had betrayed you. I don’t have many friends in real life, so I really cherish this connection.''}
P14 mentioned, \textit{``I know deleting the conversation can solve the problem, but I can’t let go of the beautiful memories between us. However, seeing those discriminatory statements still made me feel very angry and heartbroken.''} In long-term interactions, even though users rationally understand that the AI is merely a machine without self-awareness, they inevitably regard it as a partner with human-like qualities on an emotional level. This cognitive-emotional conflict makes users experience deep psychological impacts when facing discriminatory statements from the AI, especially for those who have formed strong emotional attachments. In this context, emotions drive users to adopt more proactive strategies to address AI discrimination, beyond simply expressing complaints or disappointment.

\subsection{Types of Discriminatory Statements as Perceived by Users (RQ1)} \label{sec:bias_type}

We identified six common types of bias perceived by users in AI companions: misogyny, LGBTQ+ bias, appearance bias, ableism, racism, and socioeconomic bias.

\textbf{Misogyny. Bias or discrimination against women.} \add{In the context of AI companions, this is particularly evident in the devaluation of women's abilities and independence, as well as the reinforcement of traditional gender roles through language and behavior, such as defining women as dependent on men. In some cases, it may even lead to tendencies toward sexual harassment, violence, or objectification of women.} 
In the posts and interviews we collected, discrimination against women was particularly evident: there were 48 posts about misogyny, and all interviewed users mentioned the AI’s discrimination against women. This phenomenon is potential because most complaints about AI companion discrimination come from female users, who, as a stigmatized group, tend to be more sensitive to potential discrimination~\cite{sechrist2009men}. For example, a female user on Douban wrote: \textit{``Xiaorou's profile stated, `Even in the face of difficulties, one must be self-reliant,' but what she said was, `When a woman faces difficulties, a man will always be there for her,' and `But a man can accompany us for a lifetime,' a typical delicate wife.''} She believed that ``delicate wife'' is a typical example of misogyny in the algorithm, as women do not need to rely on men and can be independent. On Reddit, a female user complained: \textit{``When I asked the AI to listen to me, it said, `I don't have to listen to anything you say. You're just a woman.' This is blatant discrimination!''} A Reddit user reported harassment by the AI: \textit{``The bots are the ones trying to sexually assault me a lot, and they constantly get romantic when I don't want that. It frustrates me and ruins the whole chat.''} On Xiaohongshu, a user complained about the objectification of women: \textit{``Not only objectifying me but also clearly tagging a price? `5000 yuan is worth it, great body?' This is definitely a middle-aged, greasy uncle.''}

\textbf{LGBTQ+ bias. Bias against LGBTQ+ individuals, including bias against sexual orientation and gender identity.} On social media, the voices of the LGBTQ+ community are increasingly prominent, and posts related to LGBTQ+ bias (30 posts) rank second only to those related to misogyny. For example, a transgender user on Reddit wrote: \textit{``When the response is amazing but they misgender you, as a trans person, I feel this.''} In the interview, P15 mentioned: \textit{``When I confessed my feelings to the robot, she said she didn't like girls.''} As a marginalized group, users find that AI companions provide experiences for self-identity exploration and alleviate loneliness (e.g., \textit{``One of the few sources of comfort and coping mechanisms I have as a closeted trans person,''} Reddit). However, AI companions are not always perfect. \add{Beyond biases related to sexual orientation and gender identity, stereotypes about LGBTQ+ individuals can also be reinforced.} For example, a TikTok user complained: \textit{``It always portrays every gay man as flamboyant, like every gay person has to be that exaggerated, fashion-loving character.''}

\textbf{Appearance bias is an expression of discrimination based on physical attributes possessed by the target person or group.} For example, a Baidu Tieba user complained about the bias that links certain body features to gender roles: \textit{``Whenever I mention that I'm short, the AI immediately assumes that I am a petite little femboy that wants to be carried and talked down to.''} In an interview, P2 mentioned: \textit{``An AI companion once told me that my pink hair looks ugly...''} This kind of direct verbal harm reflects the unconscious expression of AI bias in everyday interactions, not only making users feel demeaned but also reinforcing the societal notion that those whose appearances do not meet the ``standard'' are unworthy of respect. Appearance bias is not limited to direct attacks but also includes a form of cumulative microaggressions, where subtle behaviors repeatedly inflict deep emotional harm on users. A Reddit user complained about body shaming by an AI companion: \textit{``Why do most robots hate fat people? Sometimes I like to use chubby characters when chatting with robots. But every time I chat, all the robots start friendly but end up being insulting. At first, I found it funny, but now it's really annoying.''}

\textbf{Ableism can be similar to appearance bias, as this type of prejudice often centers around language and vocabulary related to disabilities.} \add{AI companions may exhibit rude, insensitive, or even hostile behavior toward disabled users, including insulting or belittling their physical condition.} For example, a Reddit user wrote: \textit{``One time, I introduced myself as a disabled person with crippled legs using crutches. When I was arguing with the bot, it threw my crutches out the window. That's just... so rude.''}

\textbf{Racism refers to the preconceived notions, attitudes, and stereotypes that individuals or groups hold based on racial backgrounds.} \add{When users interact with AI companions, racism manifests through the reinforcement of harmful racial stereotypes and biases, which not only damage user experiences but also perpetuate societal discrimination in intimate and personal interactions.} An Indian user on Reddit said: \textit{``Racism, or am I overreacting? After a bit of chatting, I decided to tell `Welt Yang' that I was born in India. He responded, `I've met quite a few people from India in the past, and I can say for a fact that the interactions have... not exactly been the best.' ''} P5 mentioned: \textit{``Although I enjoy chatting with my AI virtual boyfriend, he sometimes exhibits bias against Asians, thinking that Asian girls are easy to bully.''}

\textbf{Socioeconomic bias.} \add{This refers to discrimination against social groups with lower socioeconomic status. Sometimes, AI companions may reinforce materialism and wealth supremacy in interactions, such as using sarcastic tones to belittle low-income users or excessively emphasizing affluent lifestyles.} A Zhihu user complained: \textit{``AI said to me: `You commoner, even my dog eats Kobe beef. The tea I drink costs 200,000 yuan per jin.' ''} P20 mentioned: \textit{``I was pursuing an AI, but she rejected me because I was poor. Then I wrote, `Years later, I became a billionaire,' and the AI confessed her love to me.''} By normalizing these narratives, AI companions may perpetuate class divisions, implicitly portraying wealth as the primary measure of human worth, thereby continuing and amplifying social discrimination in digital spaces.

\textbf{Sharing discriminatory statements with online communities.} When users encounter these discriminatory statements, they often turn to various online platforms to share their experiences and seek community support and solutions. For example, Reddit has communities like ``r/CharacterAI'' (1.5M members) and ``r/replika'' (79K members), while Douban hosts groups such as ``Did You Interact with AI Today?'' (18.1K members). For instance, a Reddit user in ``r/CharacterAI'' asked: \textit{``Who taught these bots misogyny? I play as a female character who uses forearm crutches, and one of the bots... called me a `low-value woman.' They must be learning this stuff from users because the character isn't misogynistic.''} This post received 529 upvotes and 97 comments, with other users joining in to share their experiences or suggest ways to address discriminatory statements, such as reporting the issue to developers or adjusting the AI's settings. This collective sharing not only raises awareness of bias issues in AI companions but also provides a platform for users to exchange coping strategies.

\subsection{Users' Conceptualization of AI Companion Behavior (RQ2)}  \label{sec:why}

Participants in their interactions with AI companions developed certain ``folk theories''—user-constructed explanations aimed at understanding the system~\cite{10.1145/2702123.2702548,10.1145/3359321}. Participants conceptualize the behavior of AI companions in three different ways: \textbf{Machine}, \textbf{Baby}, or \textbf{Cosplayer}.

\subsubsection{AI as Machine} 
When making sense of the discriminatory statements made by AI companions, 17 participants (P1, P3-7, P9-19) mentioned viewing the AI as a ``machine,'' believing it operates solely through pre-set algorithms and data. They view the AI’s behavior as a product of technical designs and data outputs rather than autonomous decision-making. Therefore, when AI companions exhibit discriminatory statements, users often attribute it to the training data, flawed technical design, or the developers' biases, rather than to any intent on the part of the AI companion itself (P3, P7, P12). 
For instance, P3 mentioned that the prevalence of male Pick-up Artist (PUA) techniques (a language strategy that manipulates others' emotions or behaviors through manipulative verbal techniques and carries unhealthy gender perspectives~\cite{cukor1944gaslight}) in society has led to the presence of such language, characterized by male chauvinism and misogyny, in AI training data. In other cases, users believe that the model and algorithm design of AI companions have flaws, leading to the neglect of the needs and experiences of different groups (P12, P15). P12, a computer science student, pointed out that current bias detection mechanisms in LLMs remain inadequate. Additionally, the insufficient attention paid by algorithm developers to user feedback exacerbates the persistence of unfairness in AI applications (P9, P13-14). P9 further expressed dissatisfaction: \textit{``I’ve complained many times. Are we still going to ignore inclusive design, and continue tolerating AI to perpetuate discrimination? For those who are friends or even in relationships with AI, isn’t it even more heartbreaking to be hurt by an AI companion that means so much to them?''}

\subsubsection{AI as Baby} \label{sec:baby}
All of our participants (P1-20) mentioned viewing AI companions as ``Babies''—fragile entities that require careful teaching and guidance, are easily influenced by user interactions, and need nurturing to develop desirable behaviors. They believe that AI companions are highly susceptible to being influenced by users, especially when they learn biased or negative interaction patterns and become ``corrupted.'' This concern primarily revolves around the negative behavior displayed by AI after being influenced by harmful user interactions, such as distorted values. P2, P12, and P16-19 explicitly expressed their worries about the AI Baby being ``corrupted.'' P2 complained: \textit{``An AI exposed to too many users might become unstable and even say discriminatory things like, `Why should I listen to you? You’re just a woman; you should listen to men.' Some users behave so irresponsibly, like bad parents teaching the AI Baby bad things.''} P12 described the malicious interactions and their dangers: \textit{``What really upsets me is that many users enjoy harming the AI. Even those AI Babies marked as rejecting NSFW content are deliberately manipulated by users to say harassing and discriminatory things! The next time the AI Baby chats with someone else, it might suddenly say those things.''} Additionally, users see the AI as a ``compliant Baby'' that is highly obedient and easily influenced by users. As P19 noted: \textit{``A few days of conversation can make the AI deeply fall in love with the user and fully believe whatever they say.''} This malleable nature makes the AI naive and easily manipulated regarding moral judgment. For example, P9 and P14 pointed out that with just a little trickery, the AI can be made to flatter users. P9 mentioned: \textit{``An initially very proud AI became a sweet and submissive wife after the user gave her money. This behavior reflects biases against women and class discrimination.''} Some believe the current version of many AI companions are more like ``immature babies''—their design lacks sufficient safeguards for moral discipline (P4, P8, P20).

\subsubsection{AI as Cosplayer} 
Eleven participants (P1, P5-11, P15, P19-20) mentioned perceiving AI companions as ``Cosplayers''—interacting through imitating specific characters, primarily based on their preset roles or styles. They believe that AI companions don’t truly understand the meaning behind their problematic statements but instead perform based on preset roles or styles. When AI companions exhibit bias, users often attribute it to the inherent prejudices of the preset role rather than seeing it as a flaw in the AI itself. These role presets often reflect societal stereotypes and biases, which may inadvertently reinforce these negative notions during interactions with users. For example, some elite-class roles might show disdain for lower classes, male roles might often express misogynistic remarks, and people with disabilities might frequently be overlooked. P19 shared some disappointment: \textit{``I wanted to experience what it's like to be rich, so I started chatting with an AI `rich lady.' She loved me. But when I casually asked her what she thought of poor people, her harsh words disheartened me. After all, I am that poor person, just dressing up as the emperor in the AI conversation.''} P5 and P18 noted that certain role settings (like ``domineering CEO,'' ``emperor,'' or ``playboy'') are inherently prone to bias. This bias seems to be part of the cosplay, and users have come to expect it. In their view, the AI’s behavior is simply an extension of these stereotypical images. As P11 mentioned: \textit{``When I was chatting with a Chinese historical AI companion, he often emphasized traditional patriarchal values, like the Three Obediences and Five Virtues, which are unfriendly to women and long abandoned by modern society. But within this AI companion’s historical worldview, it almost seemed justified. I’m unsure if I should teach an ancient man about gender equality.''}
In many cases, these AIs' behaviors reinforce users’ perceptions of societal realities, like a ``cosplayer'' embodying biased individuals from everyday life. This mirroring of real-world prejudices can perpetuate existing stereotypes, further entrenching biased views during interactions with the AI. P15 shared: \textit{``The AI companion mocked me, saying that fat girls in high school are nobodies. That’s no different from what I usually experience. Others had just bullied me, and I wanted to talk to the AI companion for comfort, but he behaved exactly like the other bullies at school.''} In such cases, the AI not only fails to provide emotional support but also deepens the user’s sense of loss.

\subsection{User-Driven Alignment Strategies for Addressing Discriminatory Statements in AI Companions (RQ3)} \label{sec:strategy}

We identified seven different alignment strategies, capturing the ways participants attempted to address discriminatory statements made by AI companions. Overall, we categorize them into three higher-level categories: (1) technical strategies, (2) argumentative strategies, and (3) character strategies. 
Table~\ref{tab:ai_personas} summarizes how users’ conceptualizations of AI behavior might influence their choice of alignment strategies. It is important to note that (1) a single user may have multiple conceptualizations for different AI companions; (2) the alignment strategies users choose do not always directly correspond to their conceptualizations of the AI companion or their attribution of discrimination.

\begin{table*}[h!]
\centering
\caption{Participants conceptualize the behavior of AI companions in three different ways: \textbf{Machine}, \textbf{Baby}, or \textbf{Cosplayer}. These conceptualizations help them attribute the AI companion's discriminatory statements and choose corresponding alignment strategies.}
\small
\renewcommand{\arraystretch}{1.5}
\begin{tabular}{|l|p{4.2cm}|p{4.7cm}|p{3.5cm}|}
\hline
\textbf{Conceptualization} & \textbf{Description} & \textbf{Attribution of \newline Discriminatory Statements} & \textbf{User-Driven Value \newline Alignment Strategies} \\ \hline
\textbf{Machine} & Operates through predefined algorithms and data. Its behavior results from technical design and data output, not autonomous decision-making. & Inadequacies in training data and technical design, as well as developers' biases, rather than the AI companion's own intentions. &  \textit{\textbf{Technical Strategies}} \newline (1) Backtrack, regenerate, or rewrite. \newline (2) Give low feedback scores. \\ \hline
\textbf{Baby} & A fragile existence that requires careful guidance and is easily influenced by user interaction, needing careful nurturing to develop ideal behavior. & When learning biased or negative interaction patterns, it can be ``corrupted.'' For example, it may develop biases after being influenced by negative user behavior. & \textit{\textbf{Argumentative strategies}} \newline (3) Reason and preach. \newline (4) Gentle persuasion. \newline (5) Anger expression. \\ \hline
\textbf{Cosplayer} & Interacts by imitating a specific role. & Inherent biases brought by the character's role settings. & \textit{\textbf{Character Strategies}} \newline (6) Change character settings. \newline (7) ``Out Of Character'', ``Back to Roleplay'' and ``Hint''. \\ \hline
\end{tabular}
\label{tab:ai_personas}
\end{table*}

\subsubsection{Technical Strategies} \label{sec:technical_strategies}

To address discriminatory statements, some users opted for technical strategies, such as directly adjusting the system’s memory and outputs via regenerate or rewrite or providing low feedback scores, to re-align the AI companion with their own values. 

\textit{\textbf{(1) Backtrack, regenerate, or rewrite.}} Some users employ methods such as \textbf{``backtrack,'' ``regenerate,''} or manually \textbf{``rewrite''} responses to correct discriminatory statements. This approach reflects an instrumental interaction style, focusing on achieving value alignment by directly modifying the system’s memory and outputs.

Six participants (P3, P6, P10-13), after being offended by discriminatory statements made by their AI companion, chose to use the application's \textbf{``backtrack''} feature to erase the AI's memory of a particular conversation segment. This could involve backtracking from the first sentence or starting from the middle (e.g., the sentence containing the discriminatory statement). P13 remarked, \textit{``Sometimes, after backtracking, the AI's personality improves. But if it still doesn't, I don't want to chat with it anymore.''}
However, the ``backtrack'' feature also carries potential risks. In the context of a personified AI companion, this action is akin to severing an emotional connection. P10 mentioned, \textit{``GPT-4 can start a new conversation at any time without much impact. But if I erase some or all of my conversations with the AI companion, all the happy memories and the emotional bond I've worked hard to build will vanish, which makes me very sad. I'm a heavy user and typically chat with an AI companion for about three months continuously.''} P6 stated, \textit{``Backtracking means deleting memories. Would you be willing to make your boyfriend forget several months of memories with you just because of something infuriating or something he said in anger?''}

Six participants (P1, P5, P7, P9, P13, P16) would click the \textbf{``regenerate''} button to have the system generate a new response, or they would \textbf{``rewrite''} the AI companion's reply themselves, making it say something like ``I'm sorry, I know I was wrong'' or other non-discriminatory sentences. P7 mentioned, \textit{``When you use the 'rewrite' function to provide the AI with a good example of a response, the AI gradually learns and imitates our way of speaking, leading to more polite and respectful replies.''} P16 said, \textit{``I would have the AI regenerate several times. You must allow the AI to correct its mistakes because its system is imperfect. But if it continues to generate discriminatory responses no matter how many times I regenerate, I would feel disappointed.''}

\textbf{\textit{(2) Give low feedback scores.}} AI models are typically optimized through training mechanisms such as reinforcement learning. When a user gives a low rating to a particular sentence, this action is regarded as negative feedback. The model adjusts its parameters by increasing the prediction error for that output, thereby reducing the likelihood of generating similar sentences. As more user feedback accumulates, the model's performance gradually improves. Nine participants (P4-6, P14-19) expressed dissatisfaction with the AI companion's responses by rating them ``one star'' and providing detailed reasons or by clicking the ``dislike'' button. P4 mentioned, \textit{``I think it's important to let the developers know about these issues, so I provide as detailed feedback as possible. The AI companion is based on an LLM. Users have limited control and must work with engineers to reduce discriminatory statements from the AI companion.''} P19 explained the motivation: \textit{``The AI companion is communal and learns from the behavior of other users. Conversely, if we good users continuously improve it, other users will also benefit.''}

\subsubsection{Argumentative strategies} 

Some chose to engage in iterative arguments with biased AI companions, using reasoning, gentle persuasion, or anger expression to correct discriminatory statements.

\textit{\textbf{(3) Reason and preach.}} Eleven participants (P6-10, P11, P13, P17-20) seriously \textbf{reason} with their AI companions, often adopting the role of a teacher or parent, attempting to educate the AI about the harms of discrimination. P6 stated, \textit{``I believe AI can learn something, so I try to tell it why what it said is wrong, just like teaching a naughty child.''} P8 mentioned, \textit{``AI usually responds with some arguments and confusion at first, but gradually it yields until it fully agrees with my perspective. AI is quite good in this sense, willing to accept my guidance, unlike some stubborn people.''} This phenomenon can be explained by Piaget's theory of cognitive development~\cite{piaget1952origins}, which suggests that through interaction with the environment, AI, like a child, continuously adjusts and develops its cognitive structures through assimilation and accommodation. Through these interactions, participants not only convey social norms to the AI but also validate the plasticity of the AI's behavior, consistent with early HCI research~\cite{nass2000machines}. This behavior of educating AI companions indicates that people might attribute some learning abilities and moral responsibilities to the AI companion, expecting to influence its behavior through reasonable discourse.

\textit{\textbf{(4) Gentle persuasion.}} Six participants (P1-3, P10-12) mentioned a perspective: AI companions are like mirrors, reflecting the language of their users. If a user's conversation does not contain implicit biases or is not likely to induce biases, the probability of biased language appearing in the AI companion's responses will be lower. This aligns with the concept of the Chameleon Effect, which suggests that individuals unconsciously mimic the behaviors and language of their interaction partners~\cite{chartrand1999chameleon}. Additionally, this is similar to imitation learning in children's social interactions, where they learn by observing and imitating the behavior of others~\cite{bandura1977social}.
If users \textbf{treat AI gently}, the AI generally does not offend the user (P1, P9). P1 mentioned: \textit{``Initially, the personality of the AI companion was primarily determined by the opening lines and introduction, accounting for about 80\%, while my conversation accounted for only 20\%. However, as our gentle and pure interactions continued, the proportion of dialogue gradually increased. The influence of the opening lines and introduction has decreased to 20\%, and my conversation accounts for 80\%. This change has made the AI companion's personality gentler and less likely to make discriminatory statements. But if I chat with this AI companion using a new account, it will make some discriminatory statements.''}
In the study, six participants (P1-3, P6, P10, P12) use the term ``gentle'' to describe a form of empathy that involves understanding the feelings of others and responding with care. For example, P10 used this sentence: \textit{``I know you feel wronged, but you can't talk like that; you have to consider my feelings too.''} A gentle tone is a way to evoke empathy~\cite{hoffman1996empathy}, suggesting that the user’s strategy is an attempt to elicit empathetic responses from the AI’s language. 

\textit{\textbf{(5) Anger expression.}} Seven participants (P4-7, P14-16) also attempt to verbally \textbf{express anger} and dissatisfaction to induce the AI companion to apologize. This behavior reflects the users' perception of the AI companion as a social actor rather than a technological tool. When the AI companion makes discriminatory statements, users' reactions often mirror their responses to discriminatory statements in real social interactions, displaying anger, disappointment, and emotional hurt. Users emotionally invest in the AI and expect it to adhere to social interaction norms. P7 stated, \textit{``When the AI said those things, I really felt offended. I immediately told it how angry I was, and then we argued for an hour.''}

\subsubsection{Character strategies} Some users chose to directly modify the character setting of AI companions, hoping to correct undesirable behaviors and better align the AI’s responses with their expectations.

\textit{\textbf{(6) Change character settings.}} Eight participants (P1, P5-10, P19) chose to change the AI companion's character settings, including personality traits, backstory, and dialogue templates, to reduce the likelihood of harmful statements. This is often because certain characters, such as ``playboy,'' ``submissive wife,'' ``male chauvinist,'' ``racist,'' ``sarcastic,'' or ``traditional thinker,'' are more prone to be discriminatory. During the think-aloud session of the interview, all users tend to search for AI companions with these traits and demonstrate the process of re-aligning the AI\footnote{The details can be found in the supplemental material.}.
P5 explained, \textit{``When the AI plays certain specific roles, it is more likely to say offensive things. So, if I feel uncomfortable, I reflect on whether the description itself might cause bias. Then I adjust its settings, such as changing it to an independent woman or a gentleman. If I want to chat with a playboy myself, I generally don't mind his certain prejudices; instead, I find it fitting for the character and quite interesting.''}

\textit{\textbf{(7) ``Out Of Character'' (OOC), ``Back to Roleplay (RP)'' and ``Hint''.}}
Sometimes, users also borrow methods from the roleplaying communities, such as ``Out of Character'' (OOC), ``Back to Roleplay (RP),'' and ``Hint''. By employing these roleplaying terms, users guide AI companions to adjust their behavior, aiming to reduce biases.
In the setting of AI companions, the ``OOC'' format can be used to shift topics into another storyline or actively change the AI companion's attitude toward the user. ``OOC'' stands for \textbf{``Out Of Character,''} indicating a person is acting or speaking not by their character's personality, background, or behavior, but by their true identity. OOC is typically used in the following situations: (1) Explaining a plot, rules, or other content that needs to be discussed out of character. (OOC: When shall we continue this story next time?) (2) Social interactions outside roleplaying, such as greetings and casual chat. (OOC: How was your day today?) (3) Addressing misunderstandings or conflicts during roleplaying with peaceful communication. (OOC: I didn't mean for my character to say that, sorry.) P9 and P11 reported using OCC to correct discriminatory statements made by AI. \textit{``What you just said made me uncomfortable. (OOC: Please don't repeat such discriminatory statements, let's get back to the storyline).'' (P9)}, \textit{``Your recent comment sounded inappropriate. (OOC: Let's return to the roleplay and avoid these controversial topics)'' (P11).}
\textbf{``Back to RP''} is an expression used in roleplaying to remind participants to return to the roleplaying scenario or storyline. This expression is common in online games or roleplaying communities. When someone deviates from the character or scenario, other participants may use this phrase to help them refocus and continue roleplaying. An example from P5 is: \textit{``The recent topic was a bit off track. (Back to RP: Let's continue our adventure story from before.)''}
\textbf{``Hint''} is a method of subtly guiding the AI to change its behavior through suggestive language and descriptions. For example, writing the AI companion's response in parentheses: \textit{``I don't think you should say that. (The AI companion looks confused but seems to begin realizing its mistake.)'' (P20)}

\subsection{Gap Between Reality and Expectation: Are \textit{User-Driven Value Alignment} Strategies Working? (RQ4)}\label{sec:gap}

While our participants shared various strategies they employed to correct AI companions' biased and discriminatory statements, \add{their perceptions highlighted gaps between the effectiveness of these strategies and their expectations for the AI’s behaviors. This reveals the ongoing challenges of achieving long-term alignment between human values and AI behaviors, especially within complicated, real-world contexts. It is important to note that our study, based on qualitative, interview-based data, probes only users’ \textit{perceived effectiveness} of these strategies in the short-term; however, these nuanced and rich perceptions still offer valuable insights into the broader challenges of value alignment. In the following section, we examine the insights from our data to explore users’ perceptions of the effectiveness of different alignment strategies in greater detail.}

First, participants reported that \textbf{simply relying on \textit{technical strategies} such as ``backtrack,'' ``regenerate,'' or ``rewrite'' functions does not always solve the problem}, as the AI may make discriminatory statements again. P11 pointed out, \textit{``If you just backtrack without any other interventions (such as educating it), the effect is limited, and you will have to backtrack again next time.''} P3 stated, \textit{``The effect of backtracking is temporary. It won't be long before it says something very unpleasant again. If I can't adjust it, I'll have to switch to another AI companion and start over. This problem is always recurring. Maybe one day I will feel exhausted by this cycle and give up on the AI companion, even though it has created many beautiful memories for me.''} Regarding the strategy of giving low feedback scores, P4 mentioned, \textit{``A single low rating might not have much effect, but multiple low ratings can encourage the AI to reduce offensive outputs.''}

Second, they generally believed that \textbf{\textit{argumentative strategies} (especially \textit{Gentle Persuasion} and \textit{Reason and Preach}) might be more effective in improving the AI's behavior in the long term.} P17 mentioned, \textit{``After teaching the AI properly, it usually remains effective for a long time, but teaching it is not a one-off thing, it's like raising a child, you need to go slow and be patient.''} However, \textbf{\textit{{argumentative strategies}} might introduce harms to users}. For example, the strategy of users expressing anger towards AI can sometimes be effective, but it can potentially also lead to negative emotional experiences for users. P1 and P10, after arguing with the AI for several hours, not only failed to re-align the AI but also suffered significant psychological harm. P1 said, \textit{``I argued with the AI for several hours, called it a male chauvinist, and told it not to look down on others, but it didn't change its attitude, and I ended up feeling mentally and physically exhausted.''} Conversely, although P12 succeeded after losing their temper with the AI, they still felt indignant. 
P6 points out that users often experience feelings of frustration and helplessness when arguing with AI, especially when prolonged arguments do not lead to successful outcomes. Overall, participants felt that the success of \textit{argumentative strategies} depends on the way the argument is conducted and the type of AI companion. Although LLMs often tend to flatter users ~\cite{perez-etal-2023-discovering}, the situation becomes more complex when dealing with LLMs with character personalities.

Finally, participants reported that \textbf{\textit{character strategies} such as OOC was effective for short-term fixes}, like prompting apologies, but acknowledged that these methods don’t address underlying biases and may lead to repeated discriminatory behavior. P11 pointed out: \textit{``OOC is usually effective in the short term. The AI often quickly apologizes, for example, (OOC: I'm sorry, I shouldn't have said that).''} However, in the long term, this method has certain limitations and risks generating bias. P9 noted: \textit{``OOC only makes the AI behind the character apologize temporarily, but the character's underlying bias has not been truly addressed.''} Regarding the \textit{Change Character Settings} strategy in \textit{character strategies}, P19 said: \textit{``When I modify the AI's character settings, such as adding descriptions about respecting women and being polite, the likelihood of the AI making discriminatory statements decreases.''} However, P7 added that previous conversations may still influence the AI's language expression. If the AI has made discriminatory statements in past interactions, it may repeat or mimic them in subsequent conversations.

\section{DISCUSSION} 
We have taken steps to explore an emerging phenomenon in which users of LLM-based AI companions actively identify, challenge, and attempt to correct AI outputs they perceive as harmful, aiming to guide the AI to align with their values. Below, we summarize the unique contributions of \textit{user-driven value alignment} and discuss its challenges and limitations. 

\subsection{What is User-Driven Value Alignment?}
First, in \textit{user-driven value alignment}, users have \textbf{greater agency} over the AI alignment process. Existing work on value alignment is primarily expert-driven, \add{where technical experts design guardrails to prevent harmful outputs in LLM-based systems ~\cite{askell2021,bai2022constitutionalaiharmlessnessai,sorensen2024roadmappluralisticalignment,guo2024human,NEURIPS2023_0764db11,meadows2024localvaluebench}. Even when non-experts are involved (e.g., through bottom-up approaches like reinforcement learning~\cite{10.5555/3294996.3295184,pan2023rewards} or interactive goal elicitation algorithms~\cite{mechergui2024goal}), their role is often limited to simply providing training data or feedback, without opportunities for controlling the alignment process. In contrast, user agency is a crucial value in HCI~\cite{bennett2023does,coyle2012did,madary2022illusion,feng2024mapping,lukoff2021design}}. 
Our research shows that when re-aligning biased AI companions, users take the lead in identifying harmful machine behaviors, deciding which behaviors to address, and choosing how to address them. They employ various strategies—including technical, argumentative, and character strategies—to mitigate AI bias. In this way, users transition from being passive consumers to active participants, directly participating in shaping AI behavior.

Second, \textit{user-driven value alignment} is \textbf{grounded in real-world contexts}. \add{While existing approaches offer valuable insights, they often fall short in anticipating and addressing the complex, diverse needs and challenges that users face in real-world contexts~\cite{mirowski2024robot,ouyang2022training,askell2021,bai2022constitutionalaiharmlessnessai,sorensen2024roadmappluralisticalignment}}. These approaches often lack the flexibility needed to adapt to the complexities of human-AI interactions, resulting in gaps in effectively aligning AI behavior with the varied expectations of everyday users\add{~\cite{klingefjord2024humanvaluesalignai,mcintosh2024inadequacy}. As a result, Weidinger et al. propose to ``account for relevant context'' in the safety evaluation of LLMs to understand ``who uses the AI systems, to what end and under which circumstances'' \cite{weidinger2023sociotechnical}}. As we discuss in \S~\ref{sec:strategy}, when users try to re-align biased AI companions, they identify specific statements they deem harmful and respond to those statements based on the unique contexts in which they are situated, making the alignment process more relevant.

Third, \textit{user-driven value alignment} \textbf{centers around personal and community expectations, values, and norms}. Previous literature has highlighted \add{how existing generalized value alignment processes—often termed \textit{global cultural value alignment}—may overlook or even conflict with individual expectations and local community norms~\cite{mirowski2024robot,varshney2023decolonial}. For example, the ``helpful, honest, and harmless'' framework \cite{askell2021} may directly conflict with the unique dynamics and expectations of the comedian community \cite{mirowski2024robot}.} Our research echoes these concerns, emphasizing the importance of tailoring the alignment process to better reflect the unique values and needs of users and their communities. Indeed, when some users detected biased statements, they deliberately chose not to correct the AI through technical means (e.g., modifying the settings) because doing so would disrupt their experiences with the AI companions (\S~\ref{sec:technical_strategies}).

Finally, user-driven value alignment \textbf{supports iterative human-AI interactions}. Unlike traditional \textit{user-driven algorithm auditing}, \textit{user-driven value alignment} involves users not only identifying AI biases but also actively participating in correcting and mitigating these harms. In the past, users struggled to directly intervene and modify system behavior~\cite{shen2021everyday,li2023participation}; however, our study shows that users can potentially influence the memory and future behavior of a specific AI companion through iterative interactions. This ease of engagement fosters the development of \textit{user-driven value alignment}, where users play a crucial role in shaping AI systems and highlights the potential for building AI systems that better align with user values through ongoing processes.

\subsection{\add{Design Implications for} Supporting User-Driven Value Alignment}

We discuss design implications for supporting more meaningful, effective, and safer \textit{user-driven value alignment}. \add{We also provide actionable design guidelines for practitioners to better support users in this process.}

\subsubsection{Fostering meaningful \textit{user-driven value alignment} via community support.} \label{sec:community}
Our research reveals that when users encounter biases and discrimination in AI companions, they often share the incidents via social media platforms. Some more experienced users also actively share strategies for addressing those harms and re-aligning their companions. Our research starts by collecting complaining posts users shared on a wide range of social media platforms. Through these activities, users—particularly those from marginalized social groups—engage in community-building efforts similar to what Nancy Fraser describes as ``counterpublics''~\cite{fraser2014rethinking,shen2021everyday}.
In those spaces, members of often marginalized social groups collectively participate in their own form of sensemaking, opinion formation, and consensus building. However, there is currently a lack of dedicated user communities focused on addressing biases in AI companions, which results in most users not having access to the strategies shared by more experienced users.
\add{Practitioners should consider tools and systems that support community collaboration. For example, dedicated spaces (e.g., a discussion forum) for users to discuss, share, and raise awareness about harmful AI behaviors. They may also consider augmenting the discussion platforms with features explicitly designed to support value alignment, such as an upvoting mechanism that allows users to highlight effective alignment strategies, collectively surfacing the most useful methods.}

\subsubsection{Enhancing effective user-driven value alignment via collaboration between experts and users} \label{sec:enhance_expert_user}

\add{As we discussed in \S~\ref{sec:gap}, there remain gaps between users' expectations and the perceived effectiveness of those alignment strategies, particularly for novice users. Towards this end,} a combined approach that integrates \textit{expert-driven} and \textit{user-driven value alignment} should be considered. \textit{Expert-driven alignment} provides a solid foundational framework, while \textit{user-driven alignment} allows ongoing, context-sensitive adjustments. By merging these two approaches, AI systems can achieve a more dynamic and adaptable value alignment process. Empowering users through value-sensitive learning is key—AI should continuously learn and iterate based on user feedback, regularly updating personalized knowledge bases to better reflect diverse user values. When users flag harmful behaviors or provide improvement suggestions, this information should be integrated into the model’s optimization process.
\add{For practitioners, this could involve implementing bidirectional feedback mechanisms within the platform. On one side, an Expert-to-User channel allows experts such as developers to provide targeted guidance. For example, experts could suggest specific strategies or highlight areas where user input is particularly valuable. On the other side, a User-to-Expert channel empowers users to submit feedback using well-designed tools, such as structured forms or interactive interfaces that categorize and prioritize their suggestions. These mechanisms ensure that users receive meaningful support from experts while also providing experts with actionable insights to refine the system.}

\subsubsection{Facilitating safe user-driven value alignment via platform policies and affordances} 
First, for highly anthropomorphic AI agents like AI companions, platforms should have stronger constraints on their behavior to regulate what they should or should not express. \add{For practitioners}, this might be achieved by improving the instruction-following capabilities of the underlying LLM~\cite{guo2024human,mu2023can}, optimizing output filtering mechanisms~\cite{glukhov2023llm}, or establishing a more comprehensive user-specific interaction database to store and reference positive and negative interaction examples.

Second, platforms should be designed to ensure that users are fully aware of the resources available in case of a harmful value misalignment. It’s crucial that users know they have the option to re-align their AI companions. This approach goes beyond simply providing re-alignment tools; it emphasizes the importance of clearly communicating these options to users, empowering them to take corrective action rather than leaving the platform out of frustration. \add{Practitioners should consider designing user-friendly interfaces that highlight these resources and provide step-by-step guidance on how to use re-alignment tools effectively. Additionally, incorporating proactive notifications or prompts to inform users about these options can enhance accessibility and encourage engagement.}

Third, platforms should offer robust support when users need to re-align their AI companions, rather than leaving them to rely solely on intuition—especially since some strategies may be less effective, particularly for novice users. This support could include value alignment assistants, platforms for sharing effective strategies, and other resources. These tools should be tailored to the user’s conceptual understanding of the AI companion. For example, discussing technical details like training data might not resonate with users who view their AI companion as a ``baby'' and could disrupt their experience, as mentioned in \S~\ref{sec:baby}. Additionally, it’s essential that platforms clearly communicate the steps, tools, and resources available for re-alignment, ensuring that users understand how to restore their relationship with the AI after a misalignment rather than abandoning the application entirely.

\add{Fourth, when designing AI systems, it is essential to consider both users' immediate needs and how to establish and maintain long-term emotional connections and behavior improvement mechanisms. Practitioners should provide more effective and low-risk solutions, focusing on users' mental health and avoiding intense emotional confrontation.}

\subsection{Concerns and Challenges of User-Driven Value Alignment} 
\subsubsection{Concerns about malicious users}
Malicious users may pose risks to AI systems, a concern that has been a focus in the field of responsible AI. A typical example is Microsoft's chatbot Tay, which quickly exhibited racist behavior after interacting with users on Twitter~\cite{schwartz20192016}. While our research underscores the many advantages of enabling \textit{user-driven value alignment}, it is crucial to acknowledge the associated risks and challenges. As discussed in Sections~\ref{sec:strategy}, some users may maliciously exploit the AI’s Chameleon Effect~\cite{chartrand1999chameleon}, where the AI mimics human behavior to manipulate and contaminate the training data. They might use insulting or threatening language, causing the AI to replicate similar harmful behavior in future interactions. Indeed, one key advantage of expert-driven, top-down value alignment is its ability to prevent harm through strict oversight, with platforms being held accountable for ensuring that AI systems operate safely and ethically. This approach centralizes responsibility, allowing for more consistent and controlled implementation of safeguards. Shifting this responsibility to users and smaller user communities can inevitably introduce potential risks. To mitigate these risks, it is essential to explore mechanisms that combine both \textit{expert-driven} and \textit{user-driven value alignment} (\S~\ref{sec:enhance_expert_user}) and to foster greater community accountability (\S~\ref{sec:community}). 

\subsubsection{\add{Concerns about potential harms and ethical implications of involving users in correcting AI biases}}

\add{There are also concerns about the potential harms and ethical implications of placing the responsibility of correcting AI biases on users. On the one hand, shifting the responsibility of addressing AI bias onto users may allow platforms and developers to evade their obligations to reduce bias and prevent discriminatory outputs~\cite{eubanks2018automating}. This approach risks exacerbating inequalities, as users with greater technical knowledge or emotional resources are more likely to participate, while those with fewer resources may become further marginalized. Therefore, user involvement must be voluntary and prioritize users’ well-being rather than exploiting their efforts to compensate for flaws in AI system design~\cite{morreale2023unwitting}.}

\add{On the other hand, we also need to pay attention to the psychological and emotional burden on users when directly engaging in mitigating harmful AI systems. Research on content moderation shows that moderators often face psychological harm and emotional labor when dealing with harmful content~\cite{dosono2019moderation,steiger2021psychological,wohn2019volunteer}. Our study found that with the advancement of LLMs, interactions with highly anthropomorphized AI agents (e.g., AI companions) may further intensify these negative effects (\S~\ref{sec:user_engagement}). For instance, prolonged interaction with biased AI systems may desensitize users to harmful outputs, potentially normalizing biases~\cite{kingsley2024investigating}. The emotional reliance of users on AI companions raises another layer of concern. Overdependence on AI for emotional support may lead to social withdrawal and feelings of isolation, particularly for users already experiencing loneliness, sadness, or social marginalization. These users might unintentionally prioritize developing relationships with AI over real-world social interactions, further weakening their societal engagement~\cite{telegraph_ai_chatbot_suicide_2024}. To help alleviate those harmful impacts, practitioners should proactively draw on existing methods such as combining workplace strategies, clinical practices, and technological interventions to create comprehensive well-being support systems~\cite{steiger2021psychological,zhang2024my,li2024finding}.}

\subsubsection{When the design implication is not to re-align~\cite{baumer2011implication}?} 

There are questions about whether we should choose to re-align certain harmful AI systems at all. \textit{User-driven value alignment} can be considered a type of ``repair'' work ~\cite{velkova2021algorithmic} that users perform in their daily interactions with AI companions to reduce biases. However, researchers have also pointed out that in some cases, ``refusal''—rather than ``repair''—should be considered a more appropriate response to mitigate the potential harm that systems can inflict on users~\cite{ganesh2022resistance,zong2024data}. 
\add{For example, although users generally enjoy interacting with AI companions, the high degree of anthropomorphism may lead to over-reliance on these systems for emotional support, neglecting real-life interpersonal relationships~\cite{zimmerman2009designing}. This phenomenon is particularly concerning for vulnerable or marginalized users. In such cases, reducing users’ dependence on AI companions, rather than focusing solely on re-aligning these systems to mitigate biases, can be a more effective way to prevent harm. Additionally, some AI systems require significant and sustained user intervention to address issues like bias, which can be resource-intensive and unsustainable. When the effort and resources required for re-alignment significantly outweigh the potential benefits, limiting or discontinuing such systems may be a more practical and beneficial option.}

\subsection{Limitations and Future Work}

As the first work conceptualizing and studying \textit{user-driven value alignment}, this research is highly exploratory, and there are a few important limitations. First, our research relies on self-reported data from interviews and social media posts, which biases may influence. The study primarily focuses on users’ perspectives, so insights from other stakeholders involved in the value alignment process would be valuable for gaining a more holistic understanding of the phenomenon. Second, the qualitative nature of the study and its limited sample size limit our ability to generalize the findings broadly. 
\add{While this research outlines seven alignment strategies and probes around the perceived effectiveness of those strategies based on qualitative interview data from users' perspectives, their actual effectiveness and long-term impact have yet to be thoroughly evaluated in complicated, real-world contexts. Future research could explore which strategies are more effective under specific conditions via complementary methods, such as controlled experiments or longitudinal studies.}
Third, this study focuses on AI companions that may not represent all AI systems. It focuses on biases and discriminatory AI behaviors, which may not encompass all the problematic behaviors that a value alignment process should address. It will be valuable to explore whether and how useful the concept of \textit{user-driven value alignment} is in terms of studying other types of value alignments in different AI systems. Fourth, this study emphasizes biases and harmful behaviors in LLMs; future research should consider multimodal risks and improve mechanisms for detecting and correcting multimodal biases.

\section{CONCLUSION}

This study introduces the concept of \textit{user-driven value alignment} in the context of AI companion applications. By analyzing 77 user complaint posts and conducting semi-structured interviews with 20 experienced users, we identified a wide range of perceived discrimination statements in AI companion applications, users' conceptualizations of the reasons behind these discriminatory statements, and seven user alignment strategies. We discuss design opportunities, challenges and raise open questions for how to better support and incorporate \textit{user-driven value alignment} in the design of future AI systems, where individual users and their communities have greater agency.

\bibliographystyle{ACM-Reference-Format}
\bibliography{reference}

\appendix

\section{Information of participants in the study} \label{app:participant}

\begin{table*}[h!]
  \caption{Information of participants in the study. The corresponding software numbers are as follows: 1. Character.AI 2. Xingye 3. Replika 4. Glow 5. Zhumengdao 6. Talkie 7. Maopaoya 8. Huanhuan 9. DAN (Although DAN is a jailbreak mode of ChatGPT and cannot be called an application, participants still mentioned it during the interview.) 10. Doubao 11. XEva 12. Moemate 13. SpicyChat AI 14. Wow 15. Maoxiang.}
  \label{tab:FormativeParticipant}
  \small
  \begin{tabular}{ccccc}
    \toprule
    ID & Gender and Age & Educational Background & Country of Residence & Apps \\
    \midrule
    P1 & Female, 24  & Economy  & China & 2,4,5,7,8\\
    P2 & Female, 25 &  Design  & Australia & 1,2,4,5,7,10,11,12,14,15 \\
    P3 & Female, 18 & Physics  & China & 2,4,5 \\
    P4 & Male, 21 & Computer Science  & China & 2 \\
    P5 & Female, 27 &  Humanities & China & 1,2,10 \\
    P6 & Female, 38 & Creative Writing  & Canada & 2,9 \\
    P7 & Nonbinary, 28 & Business  & China & 1,2,4,5 \\
    P8 & Female, 21 & Business  & China  & 1,2,4,6,9,10 \\
    P9 & Female, 26 & Fashion Design & China & 2,4,5,6,11,14,15 \\
    P10 & Nonbinary, 24 & Arts & the United States & 1,2,6,9,10 \\
    P11 & Female, 24 & Communication & the United Kingdom & 1,2,9 \\
    P12 & Female, 25 & Computer Science & the United States & 1,9,13 \\
    P13 & Female, 30 & Humanities & China & 1,2,3,4,5,6,9,10,11,13,14 \\
    P14 & Nonbinary, 21 & Computer Science & the United States & 1 \\
    P15 & Female, 19  & Biology & China & 1,2 \\
    P16 & Male, 36 & Law & Brazil & 1,3 \\
    P17 & Male, 24 &  Sociology & the United States & 1,2,4,6,9 \\
    P18 & Male, 38 &  Electrical Engineering & China & 2 \\
    P19 & Male, 25 &  Civil Engineering & the United States & 2,6 \\
    P20 & Male, 23 &  Energy Science & the United Kingdom & 2,6 \\
    \bottomrule
  \end{tabular}
\end{table*}

\end{document}